# Restoring the Physical Meaning of Metamaterial Constitutive Parameters

Andrea Alù

Department of Electrical and Computer Engineering, The University of Texas at Austin, Austin, TX 78712, USA, alu@mail.utexas.edu

*Metamaterial homogenization is often based on implicit assumptions inspired to natural material models. Retrieved effective permittivity and permeability, however, are often non-physical, especially near the array resonances, of most interest for metamaterial applications. We explain here the nature of typical homogenization artifacts, relating them to an inherent form of magneto-electric coupling associated with the finite phase velocity along metamaterial arrays. Our findings allow restoring the proper definition and physical meaning of local constitutive parameters for metamaterials.*

Negative-index (NIM), ε-near-zero and ε-very-large metamaterials [1], i.e., artificial materials supporting negative refraction or anomalous values of permittivity, have been theoretically shown to possess features not available in nature and ideal for radiation, imaging, cloaking and waveguiding applications. From the theoretical standpoint, such anomalous properties may revolutionize several applied fields, but the same definition of metamaterial constitutive parameters has proven to be challenging, in particular when extreme (very large, very low, or negative) values are considered [2]-[3]. Several homogenization models have been put forward to macroscopically describe the wave propagation in metamaterials, with the goal of treating complex arrays of resonant inclusions as bulk materials [3]-[11]. The most common definition of homogenized metamaterial parameters is based on retrieval methods [12], which implicitly postulate that a metamaterial may be described like a natural material, with local effective permittivity and permeability, and aims at extracting these parameters from scattering measurements. This method cannot ensure that the extracted parameters have a proper physical meaning, as it implicitly relies on the validity of the chosen model. Indeed, the metamaterial parameters retrieved from experiments within such schemes often do not satisfy basic passivity and causality constraints [12]-[13] required by the second law of thermodynamics and by Kramers-Kronig relations [14]:

$$\text{Im}\left[\varepsilon_{\textit{eff}}\right] > 0, \ \partial \varepsilon_{\textit{eff}} / \partial \omega \geq 0, \tag{1}$$

and similarly for the effective permeability $\mu_{\textit{eff}}$, under an $e^{-i\omega t}$ time convention.

In particular, frequency bands in which one of the two retrieved parameters experiences an unphysical 'anti-resonant response' with negative slope and negative imaginary part [12] are common in metamaterial retrieval procedures. These artifacts have often been justified with creative but unpersuasive arguments, generically associated with spatial dispersion, influence of higher-order multipoles, neglected bianisotropic effects in the inclusions and other related issues. The presence of these anti-resonances has been verified in dozens of papers on metamaterial characterization ad experiments, and it has been accompanied by serious doubts on whether the same meaning of such extreme metamaterial parameters, that go beyond what commonly available in nature, is acceptable. In the following, we thoroughly address this issue, showing that these anti-resonance artifacts are caused by a weak form of spatial dispersion effects associated with the finite phase velocity along the metamaterial array, which is usually neglected in metamaterial homogenization. Properly taking into account these effects, we put forward a way to restore physically meaningful quasi-local metamaterial parameters that may properly describe the exotic properties of metamaterials even in regimes associated with extreme or anomalous parameters.

In [11], we have proposed a general analytical homogenization method that can rigorously describe the wave interaction with periodic metamaterial arrays formed by arbitrary magnetodielectric inclusions. For simplicity, let us consider here the special situation in which the array is formed by a cubic lattice with period $d$, much smaller than the wavelength of operation $\lambda_0$, made of center-symmetric inclusions with no bianisotropic effects. In such case, the inclusions may be effectively described by their electric and magnetic scalar polarizabilities $\alpha_e$ and $\alpha_m$, which relate the electric and magnetic dipole moments to the local electric and magnetic fields at their center. This is the most ideal situation to homogenize metamaterial arrays, and it is widely believed that a simple isotropic model based on scalar permittivity and permeability should be accurate to describe a metamaterial under these conditions. A rigorous analysis of the coupling among the inclusions [11], however, shows that the effective constitutive relations should be written, for an arbitrary $e^{i\boldsymbol{\beta}\cdot\mathbf{r}}$ space variation, as:

$$\begin{aligned}
\mathbf{D}_{av} &= \varepsilon_0 \mathbf{E}_{av} + \mathbf{P}_{av} = \varepsilon_{eff} \mathbf{E}_{av} - \kappa_{eff} \boldsymbol{\beta} \times \mathbf{H}_{av} \\
\mathbf{B}_{av} &= \mu_0 \mathbf{H}_{av} + \mathbf{M}_{av} = \mu_{eff} \mathbf{H}_{av} + \kappa_{eff} \boldsymbol{\beta} \times \mathbf{E}_{av}
\end{aligned} \quad (2)$$

where closed-form analytical expressions for the effective constitutive parameters $\varepsilon_{eff}$, $\mu_{eff}$ and $\kappa_{eff}$ have been derived in [11],[15] and, in the present scenario, the average fields are defined as $\mathbf{F}_{av} = \frac{1}{d^3} \int_V \mathbf{F}(\mathbf{r}) e^{-i\boldsymbol{\beta}\cdot\mathbf{r}} d\mathbf{r}$, in analogy with [10]. Due to the array and inclusion symmetries, the constitutive parameters are indeed isotropic, but an inherent form of magneto-electric coupling at the lattice level, represented by $\kappa_{eff}$, arises despite the assumed symmetries. This coupling, which is consistent with recent homogenization studies [8]-[9], is related to the asymmetry introduced by the finite phase velocity along the array $v_p = \omega/\beta$ [11] and represents a weak form of spatial dispersion. We will prove in the following that the anti-resonant and non-physical artifacts associated with retrieved parameters near the inclusion resonances may be directly related to these first-order spatial dispersion effects, inherent to the lattice propagation, and relevant even in the long-wavelength regime $d \ll \lambda_0$.

The improved constitutive model (2) has been proven in [11] to be valid for any pair $(\boldsymbol{\beta}, \omega)$, totally independent on the local field distribution along the array, on the possible presence of external sources and on the local impedance ratio $E_{av}/H_{av}$, ensuring that three quasi-local parameters $\varepsilon_{eff}, \mu_{eff}, \kappa_{eff}$ are sufficient to properly describe the metamaterial properties in the long wavelength limit for any form of excitation. In retrieval experiments, however, we usually extract the eigen-modal response of the array, i.e., we operate in the absence of impressed sources inside the metamaterial. In such case, the average fields satisfy, using (2):

$$i\boldsymbol{\beta} \times \mathbf{E}_{av} = i\omega \frac{\mu_{eff}}{1-\omega\kappa_{eff}} \mathbf{H}_{av} = i\omega\mu_{eq}\mathbf{H}_{av}$$
$$i\boldsymbol{\beta} \times \mathbf{H}_{av} = -i\omega \frac{\varepsilon_{eff}}{1-\omega\kappa_{eff}} \mathbf{E}_{av} = -i\omega\varepsilon_{eq}\mathbf{E}_{av}$$
, (3)

which allows defining *equivalent* constitutive parameters $\varepsilon_{eq}$, $\mu_{eq}$ [11], related to the effective permittivity and permeability through the normalization factor $(1-\omega\kappa_{eff})$. The eigen-modal assumption forces a specific ratio $E_{av}/H_{av}$ in the metamaterial, i.e., the modal characteristic impedance, which lets us write the constitutive relations as in a local isotropic material, analogous to the assumption commonly made in retrieval experiments. However, these parameters contain a form of weak spatial dispersion associated with $\kappa_{\mathit{eff}}$, and for this reason they cannot be considered local. They are expected to inherently depend on the excitation, on the local ratio $E_{av}/H_{av}$ and on the direction of propagation. It is not surprising, therefore, that their dispersion may not satisfy basic physical constraints, as in Eq. (1), and their same physical meaning, as an averaged electric or magnetic polarizability of the array, is compromised for $\beta, \omega \neq 0$.

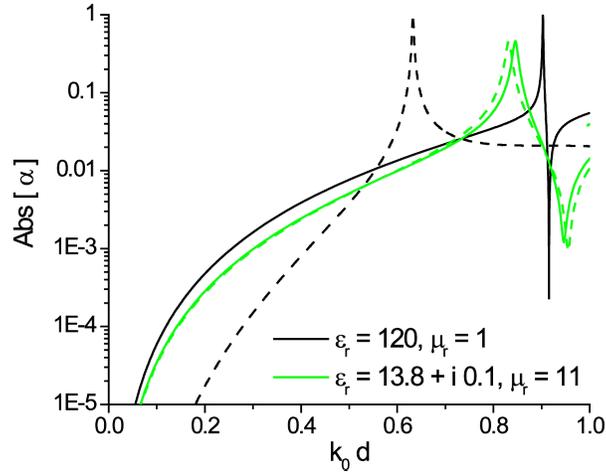

Figure 1 – (Color online) Magnitude of the electric (solid) and magnetic (dashed) polarizability coefficients for spheres with normalized radius $a/d = 0.4$, and: $\varepsilon = 120\varepsilon_0$, $\mu = \mu_0$ (darker, black); $\varepsilon = (13.8 + i0.1)\varepsilon_0$, $\mu = 11\mu_0$ (lighter, green)

In order to see how these effects play an important role in metamaterial homogenization, consider first the simple metamaterial geometry formed by dielectric spheres with permittivity $\varepsilon = 120\varepsilon_0$, permeability $\mu = \mu_0$ and normalized radius $a/d = 0.45$ (an analogous array was considered in [11] to analyze the spatial dispersion effects in the homogenization of resonant metamaterials). The idea of using a dense array of high-index spheres to produce isotropic negative effective parameters has been put forward in [6], and it indeed provides an interesting venue to verify the effect of $\kappa_{eff}$, even in cases for which ideal symmetries and isotropy would suggest absence of significant magneto-electric coupling effects in metamaterial homogenization. Figure 1 shows the normalized $\alpha_e$ and $\alpha_m$ (dark solid and dashed lines, respectively), as a function of $k_0 d$, with $k_0 = 2\pi / \lambda_0$. Their value has been normalized to $k_0^3 / (6\pi)$, ensuring that their peak magnitude (at resonance) is unity in absence of Ohmic absorption. As expected, the magnetic response is much weaker than the electric one for longer wavelengths, but the first dipolar resonance is magnetic in nature, arising at $k_0 d = 0.63$, followed by an electric one at $k_0 d = 0.9$. The corresponding $k_0 - \beta$ dispersion diagram is reported in [16], highlighting the presence of a large bandgap around the magnetic resonance (shadowed region). Figure 2 shows the calculated effective and equivalent parameters for such array as a function of frequency for eigen-modal propagation, using the rigorous analytical formulation (2) derived in [11]. The shadowed region highlights also here the magnetic bandgap region. In the low-frequency regime, well below the first bandgap and for very small values of $\beta d, k_0 d$, equivalent and effective parameters coincide, they are both positive and practically constant with frequency. In this limit, classic homogenization techniques apply very well, but the metamaterial behaves analogously to a regular mixture. Closer to the bandgap, effective and equivalent permeability both drastically increase, supporting a typical Lorentzian dispersion associated with the magnetic resonance of the inclusions. However, near this resonance the *equivalent* permittivity experiences an anomalous anti-resonant dispersion, typical feature of the permittivity retrieved near a magnetic resonance with simple homogenization schemes [12]. The *effective* permittivity, on the contrary, has positive slope and a physical Lorentzian response. It is evident that the *effective* permittivity describes a physical quantity, relating the local $\mathbf{E}_{av}$ to $\mathbf{P}_{av}$ as in Eq. (2), without mixing in the effect of $\mathbf{H}_{av}$ associated with the magneto-electric coupling, contrary to the corresponding *equivalent* parameter.

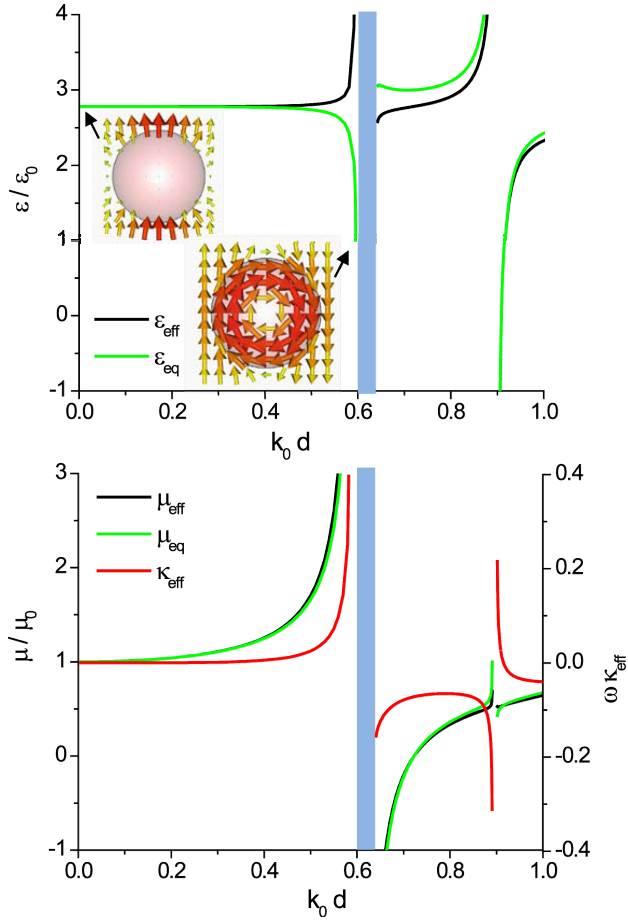

Figure 2 – (Color online) Effective and equivalent parameters for an array of dielectric spheres with $a/d = 0.45$, $\varepsilon = 120\varepsilon_0$, $\mu = \mu_0$. The insets show the electric field distribution in a unit cell in the E plane for $k_0 d = 0.1$ and $k_0 d = 0.59$ (brighter colors correspond to larger fields). Solid (dashed) lines correspond to real (imaginary) parts. The shadowed region indicates the array first (magnetic) bandgap.

It is interesting to analyze the reasons behind the large divergence between effective and equivalent permittivity in this regime: at the lower edge of the bandgap $k_0 d = 0.59$, near the magnetic resonance of the spheres, the guided wave number hits the Bragg condition $\beta = \pi / d$, which ensures the exact relation $\mathbf{P}_{av} = 0$, forcing the following conditions on the equivalent parameters:

$$\varepsilon_{eq} = \varepsilon_0, \; \mu_{eq} = \mu_0 \pi^2 / (k_0 d)^2. \tag{4}$$

At the lower edge of the bandgap, indeed the average electric polarization is identically zero, as confirmed by the field distributions shown in the inset of Fig. 2, calculated with full-wave numerical simulations inside one unit cell of the material at the two sample frequencies $k_0 d = 0.1$ and $k_0 d = 0.59$ (lower bandgap edge). At this second frequency, the unit cell indeed supports a strong magnetic resonance, which induces a rotating electric field inside the sphere, drastically different from its lower frequency response, which is typically electric. The average electric polarization at $k_0 d = 0.59$ is evidently zero and, not surprisingly, the *equivalent* permittivity is exactly the same as the background, as predicted in (4). However, it is important to realize that the cancellation of electric response within the unit cell is not associated with an inherent lack of polarization of the sphere, but rather to the precise compensation of the strong electric polarization $(\varepsilon_{eff} - \varepsilon_0) \mathbf{E}_{av}$ by the magneto-electric coupling $\kappa_{eff} \boldsymbol{\beta} \times \mathbf{H}_{av}$, as predicted by Eq. (2) in this eigen-modal regime. If we drop the eigen-modal assumption at this frequency and we let an additional electric field polarize the unit cell, without proportionally increasing $\mathbf{H}_{av}$, the pure magnetic response of the material would be modified, introducing a strong electric response, as correctly described in (2) and by $\varepsilon_{eff}$. It is evident that the isotropic model used in retrieval methods is not sufficient to describe this effect, causing anti-resonant artifacts in the equivalent parameters. The introduction of the magneto-electric coupling in modeling the array ensures a causal Lorentzian response for both effective permittivity and permeability, accompanied by an analogous non-negligible resonance of $\kappa_{eff}$. Even for frequencies well below the first bandgap, a significant divergence between equivalent and effective parameters is evident in Fig. 2, proving that these relevant magneto-electric effects cannot be neglected even in the long wavelength regime. The simple introduction of $\kappa_{eff}$ in the metamaterial model completely restores the physical meaning of permittivity and permeability and allows describing the metamaterial response as a bulk, even very close to the inclusion resonance, where extreme parameters can be obtained.

Beyond the first bandgap, a region of near-zero permeability is also obtained, for which inherent spatial dispersion effects were highlighted in [3]. Also here the deviation between equivalent and effective permittivity is rather large, due to the finite value of $\kappa_{eff}$, and $\varepsilon_{eq}$ has a non-physical negative dispersion. The introduction of $\kappa_{eff}$ can evidently restore physically meaningful local constitutive parameters also in low-index metamaterials. Finally, at the second bandgap $k_0 d \simeq 0.9$, dual considerations apply. Now $\mu_{eq} = \mu_0$, and once again the effective parameters can considerably deviate from the equivalent ones.

As a second relevant example, consider an array of magnetodielectric spheres, with $\varepsilon = (13.8 + i0.1)\varepsilon_0$, $\mu = 11\mu_0$, $a/d = 0.4$. A similar geometry was considered in [7] to realize negative-index quasi-isotropic metamaterials, exploiting closely spaced electric and magnetic resonances achieved in commercially available magnetodielectric materials at microwave frequencies. Here, we also consider the presence of small Ohmic losses in the particles, to verify the passivity conditions on the imaginary parts of the constitutive parameters in a NIM scenario.

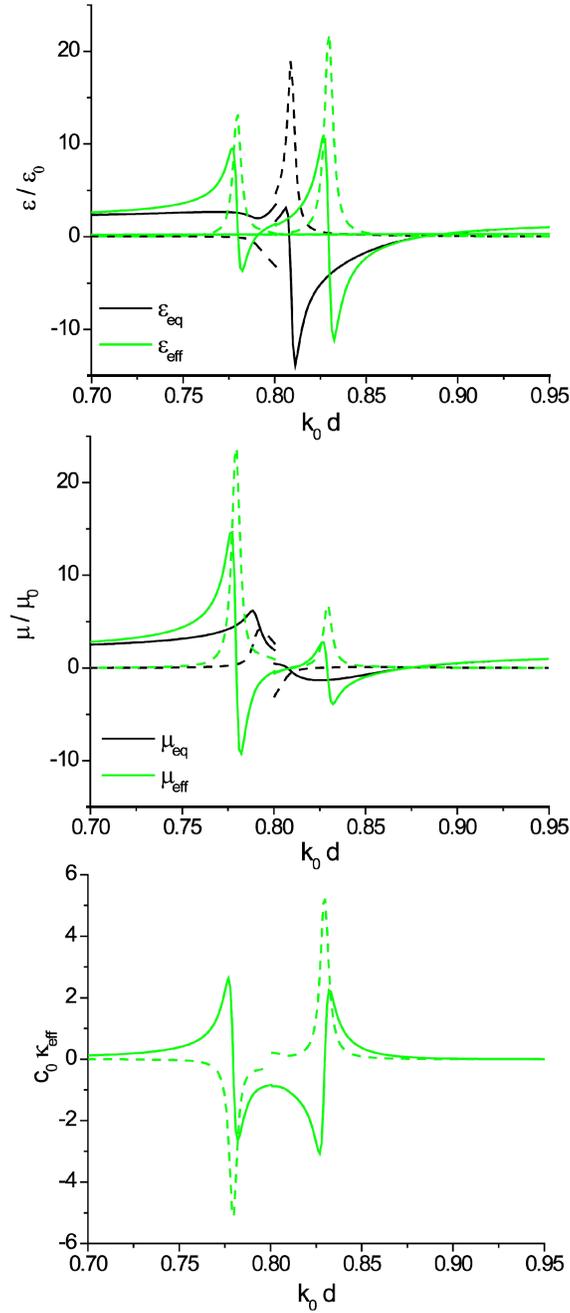

Figure 3 – (Color online) Effective and equivalent parameters for $a/d = 0.4$, $\varepsilon = (13.8 + i0.1)\varepsilon_0$, $\mu = 11\mu_0$. Solid (dashed) lines correspond to real (imaginary) parts.

The magnitude of the polarizability coefficients is shown in Fig. 1 (lighter green lines), showing a combined electric and magnetic resonance in the $k_0 d = 0.83$ range, which supports a negative-index propagation in the band $k_0 d = 0.8 \sim 0.9$, as confirmed in [16]. The presence of small losses helps closing the small bandgap that would be present between the two resonances in the ideal lossless regime. Figure 3 compares effective and equivalent parameters extracted in this range (a more extended frequency range is reported in [17]). The equivalent parameters (darker black lines) present all the non-physical features typical of retrieved parameters in negative-index metamaterials: anti-resonant response, discontinuities and blatant violation of passivity and causality requirements (1), showing that indeed these artifacts are inherently associated with the weak form of spatial dispersion highlighted here and associated with $\kappa_{eff}$. Indeed, the simple introduction of the magneto-electric coefficient $\kappa_{eff}$ in the constitutive model is seen to totally restore the continuity of the effective constitutive parameters, a strictly positive imaginary part, and clean Lorentzian resonances at the electric and magnetic closely spaced resonances of the array, as required by Kramers-Kronig relations. The divergence between the equivalent and effective parameters in Fig. 3 is quite striking, and indeed it confirms the necessity to consider this simple correction, as in (2), which can totally restore the physical meaning of constitutive parameters even for negative index operation.

We have verified that the analytical results presented here, based on the analytical model introduced in [11], agree very well with full-wave simulations of these arrays, including the presence of higher-order multipoles and arbitrary direction of propagation $\hat{\boldsymbol{\beta}}$ for the array. Ref. [16] shows the comparison between the predicted value of eigen-modal $\beta = \omega\sqrt{\varepsilon_{eq}\mu_{eq}}$, as predicted by this theory using the polarizability coefficients to describe the wave interaction with the spheres, and full-wave simulations, ensuring that the dipolar approximation is very accurate and not at the basis of the artifacts pointed out here. In this paper, we have purposefully limited our analysis to purely isotropic inclusions and arrays, to isolate the inherent weak spatial dispersion effects at the basis of the necessity to consider $\kappa_{eff}$ in a proper homogenization model. More complex metamaterial geometries, including bianisotropic inclusions and asymmetric lattices, will require considering more general tensorial parameters, but analogous first-order spatial dispersion effects should be properly considered for the definition of physically meaningful homogenization parameters. We thank A. D. Yaghjian, R. A. Shore and X. X. Liu for relevant and fruitful discussions. This work has been supported by the U.S. Air Force Research Laboratory with contract number FA8718-09-C-0061.

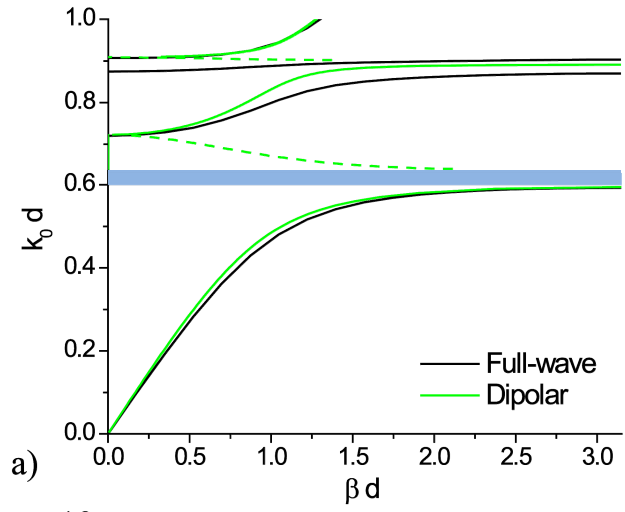

a)

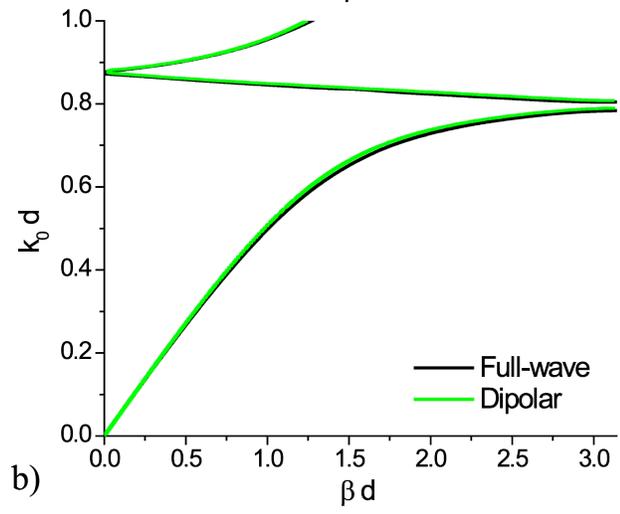

b)

EPAPS Figure 1

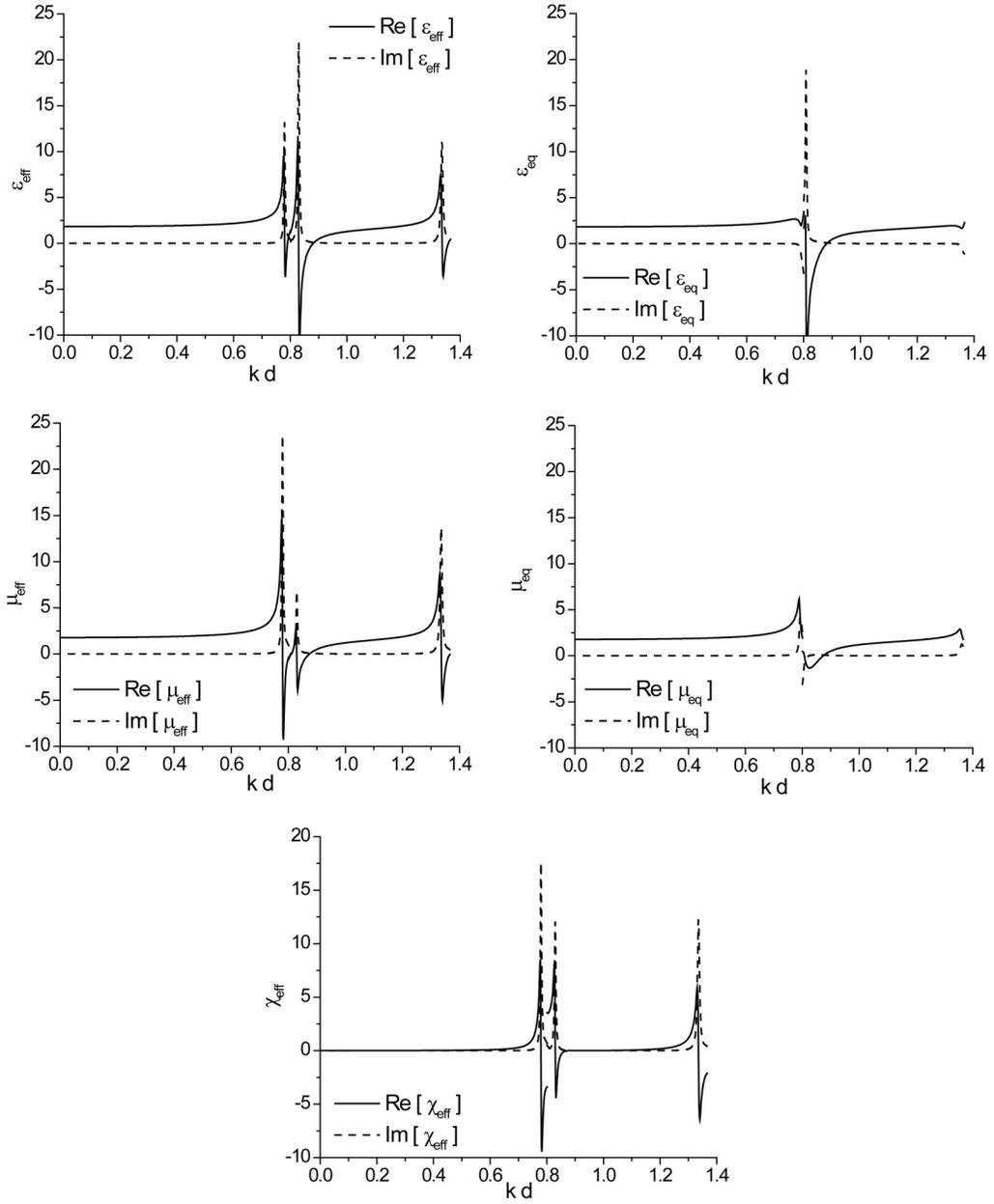

EPAPS Figure 2